\documentclass[a4paper,11pt]{article}
\usepackage{jinstpub} % for details on the use of the package, please see the JINST-author-manual
\usepackage{lineno}
\title{\boldmath Calibration of a $\Delta E-E$ telescope based on CeBr$_3$ scintillator for secondary charged particles measurements in hadron therapy }

\author[a,b]{L.~Gesson} 
\author[a]{J.~Gross\footnotemark[1]} 
\author[a]{C.~Mozzi\footnotemark[1]}
\author[a]{C.~Reibel}
\author[a]{Ch.~Finck}
\author[a]{S.~Higueret}
\author[a]{T.D.~Lê}
\author[a]{E.~Traykov}
\author[d]{J.C.~Thomas}
\author[a]{N.~Arbor}
\author[c]{M.~Pullia}
\author[a]{G.~Harmant}
\author[a]{M.~Vanstalle}

\affiliation[1]{Both authors contributed equally}
\affiliation[a]{Universite de Strasbourg, CNRS, IPHC UMR 7178, Strasbourg, 67000, France}
\affiliation[b]{GSI Helmholtzzentrum fur Schwerionenforschung GmbH, Darmstadt, 64291, Germany}
\affiliation[c]{CNAO Centro Nazionale di Adroterapia Oncologica, Pavia, 27100, Italy}
\affiliation[d]{GANIL Grand Accélérateur National d'Ions Lourds, CEA-DRF, CNRS, Caen, 14000, France}

\emailAdd{levana.gesson@iphc.cnrs.fr}

\abstract{Hadrontherapy is an established cancer treatment method that enables a more localized dose deposition compared to conventional radiotherapy, potentially reducing the dose to surrounding healthy tissues in certain clinical cases.
However, a key limitation in current treatment planning lies in the limited experimental data available for the characterization of secondary particles generated by nuclear interactions of the primary beam with tissues, which directly impacts the accuracy of Monte Carlo tools and analytical models used in dose calculations.
Indeed, this leads to the adoption of larger safety margins and can limit the use of hadrontherapy for treating certain complex or sensitive tumor locations.

This work is part of the context of the characterization of secondary charged particles generated by ion beams in the energy range relevant for particle therapy applications, using a $\Delta E-E$ telescope comprising a CeBr$_3$ crystal scintillator and a plastic scintillator. The calibration and response of this telescope to ions commonly used in clinical settings is presented in this work, highlighting adherence to Birks' law for accurate energy measurements. 
This study is the first to optimize a $\Delta E-E$ telescope combining CeBr$_3$ and plastic scintillators specifically for secondary particle detection in hadrontherapy. It represents an essential step toward the experimental acquisition of nuclear data, enabling accurate measurement and identification of secondary charged particles generated by therapeutic beams in tissue-equivalent materials. The system is designed for use in controlled experimental setups that reproduce clinical conditions, with the goal of improving the predictive accuracy of treatment planning software through enhanced Monte Carlo simulation inputs.}


\arxivnumber{1234.56789} % Only if you have one

\begin{document}
\maketitle
\flushbottom

\section{Introduction}
\label{sec:Introduction}
Particle therapy, including proton and heavy-ion therapy, has demonstrated significant advantages over conventional X-ray radiotherapy by providing highly conformal dose distributions to the tumor while better sparing surrounding healthy tissues \cite{Durante2010} \cite{Paganetti2012}.

However, the interactions between the primary particle beam and human tissue lead to beam and target fragmentation, producing secondary particles such as lighter ions, neutrons, and gamma rays. These secondary particles can contribute to out-of-field dose deposition through ionization and nuclear interactions in surrounding tissues. This damage can affect cellular functions and lead to biological responses that can contribute to undesired dose in healthy organs and tissues \cite{Schardt2010} \cite{Bohlen2014}. This additional dose is not accounted for by the primary beam alone and must be carefully modeled to ensure accurate treatment planning and risk assessment.

Accurate dose calculations for particle therapy rely primarily on analytical or semi-empirical models integrated in treatment planning systems, that model complex physical and biological processes. These models are often supported or benchmarked using data from detailed Monte Carlo simulations such as Geant4 \cite{Sarrut2014} \cite{Agostinelli2003}. However, there is currently a lack of experimental data regarding nuclear reactions in particle therapy, which can introduce inaccuracies in dose calculations \cite{Kox1987}\cite{DiRuzza2022}, potentially affecting clinical treatment margins and robustness, as highlighted in \cite{Paganetti_2012}.

The CLINM (Cross-Sections of Light Ions and Neutron Measurements) project, addresses this gap by focusing on the precise characterization of secondary particles produced during ion fragmentation in tissues, in order to compare with Monte Carlo simulations and improve their accuracy. 
%as well as their associated chemical effects on water and biomolecules,  \\

One key point of the CLINM project is to measure the yields, charges, and energies of the fragments produced under conditions that replicate clinical experimental settings, using tissue-equivalent materials. In this work, the focus will be on the charged secondary particles. To achieve this, a $\Delta E-E$ telescope detection system has been developed. This technique leverages the principles of energy loss and remaining energy to discriminate different particle charges. The thinner detector in the telescope, a plastic scintillator, measures the energy loss ($\Delta E$) of charged particles as they pass through, while a CeBr$_3$ crystal scintillator measures the remaining energy ($E$) of the particle. \\

Before employing this system for nuclear data acquisition on the production of secondary charged particles in hadrontherapy, it is essential to characterize it and assess its measurement performance. This includes energy calibration, energy resolution, ion discrimination capability in the $\Delta E-E$ mode, and time resolution for correlation studies or even time-of-flight measurements.

The presented research aims to characterize the response of the $\Delta E-E$ telescope to ions used in clinical therapy by analyzing their energy deposition patterns in the two components. Unlike previous studies, this work employs a CeBr$_3$ scintillator, a choice for heavy-ion detection in hadrontherapy applications, owing to its unique properties of high light yield, fast timing, and low intrinsic background radiation. This marks a significant improvement over traditional scintillators (e.g., NaI or LaBr$_3$) that suffer from limitations such as internal radioactivity or saturation at high energies.

In this study, the primary beams simulate the clinical energy ranges used in therapy, but the focus remains on the secondary charged particles generated during beam interactions with materials mimicking patient tissues (e.g., PMMA). This telescope is designed to measure energy deposition patterns of these secondary charged particles, which include protons, light ions (e.g., deuterons, alpha particles), and heavier fragments generated by nuclear fragmentation processes, in materials mimicking human tissues (e.g., PMMA phantoms), under experimental conditions that simulate those of clinical beams with energies ranging from a few MeV up to 200~MeV/u.

\section{Material and Methods}
\label{MatMeth}
\subsection{Detectors description}
The $\Delta E-E$ telescope is composed of two scintillating detectors: a thin plastic scintillator ($\Delta E$) and a thick crystal scintillator of CeBr$_3$ ($E$), long enough to stop $^{12}$C up to 200~MeV/u.

The CeBr$_3$ detector is a cylindrical 2$\times$2 inches crystal scintillator, from Advatech UK Ltd coupled to a R6231-100 Hamamatsu photomultiplier (PMT) \cite{Advatech}. An entrance window in front of the crystal inside the CeBr$_3$ is made of 400~$\mu$m of aluminum and 1~mm of polytetrafluoroethylene (PTFE), used as a reflector. The CeBr$_3$ detector has a short decay time of 19~ns, a high light yield of 60,000~Photons/MeV, an energy resolution of 3.8~keV at 662~keV fo X-ray and gamma-ray, and an inherent low background radiation of 0.004~Bq/cm$^3$.

The CeBr$_3$ crystal scintillator was chosen for its capacities to detect not only charged particles but also gamma rays and neutrons, aligning well with the general context of CLINM. Unlike some alternatives like LaBr$_3$, CeBr$_3$ does not exhibit internal radioactivity, which minimizes background noise and false coincidences, thus enhancing the precision of radiation detection.

Concerning the voltage applied on the R6231-100 Hamamatsu PMT coupled to the CeBr$_3$, the constructor recommended value is +~1200~V. However, the voltages applied were of +~400~V and +~350~V, in order to detect the high-energy protons and $^{12}$C without reaching the PMT saturation, following recommendations from \cite{Miller_2018}\cite{Quarati_2011}.\\

The plastic scintillator, from Eljen Technology (EJ-228), is a square of 6$\times$6$\times$0.2~cm$^3$, put in front of the CeBr$_3$. The plastic detector has a short decay time of 1.4~ns, and a light yield of 10,200~Photons/MeV. This plastic scintillator is designed to deliver high performance in terms of timing resolution and light output, making it ideal for our specific needs.

Two different photomultipliers (PMT) from Hamamatsu \cite{HamamatsuR7057}, R7057 and XP3990, were tested, coupled to the plastic scintillator, supplied with -~1100~V and -~1200~V high-voltage respectively, in order to optimize the settings, i.e. minimize saturation while maximizing resolution.\\

\begin{figure*}[!h]
\begin{center}
\includegraphics[scale=0.67]{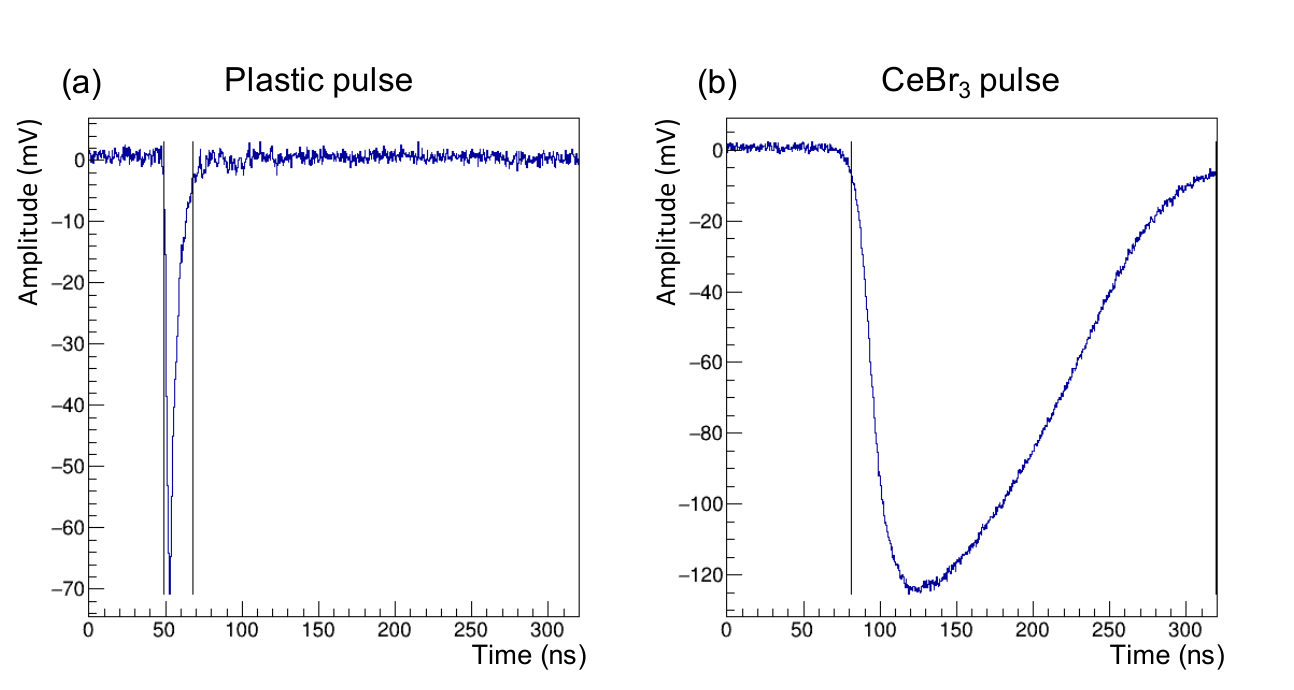}
\caption{\label{fig:pulse} Pulse shape of the signal obtained with the plastic scintillator with -~1200~V applied (a) and CeBr$_3$ crystal with +~350~V applied (b) irradiated with 200~MeV/u $^{12}$C.}
\end{center}
\end{figure*}

Both detectors signals were aquired in coincidence by a WaveCatcher digitizer module, developed by the LAL laboratory (Paris, France) \cite{Breton}. The sampling frequency was set to be 3.2~GHz on the the data acquisition system (DAQ). 
Figure \ref{fig:pulse} illustrates digitized signals obtained from the plastic scintillator and CeBr$_3$ crystal when irradiated with 200~MeV/u $^{12}$C ions. As expected, the pulse duration and the rising time of the plastic scintillator are shorter than the ones from the CeBr$_3$, in accordance with intrinsic properties of both scintillators. A gate of 80 ns was applied for the detector’s coincidence detection.

\subsection{Experimental setup}

Experiments were carried out in different facilities providing ions of various types and energies: 
\begin{itemize}
    \item \textbf{Cyrc\'e} cyclotron at IPHC (Strasbourg, France) \cite{Constanzo_2019}, producing protons beam of 25~MeV, which can be degraded by an aluminium wheel to lower energies;
    \item \textbf{CAL} (Centre Antoine Lacassagne, Nice, France) protontherapy center \cite{Hofverberg_2022}, specializes in proton therapy treatments for cancer, providing a proton beam of 62~MeV ;
    \item cave M of \textbf{GSI} (Darmstadt, Germany), the heavy ion accelerator facility SIS-18 \cite{gsi_sis18}, using PMMA thickness variations to modulate carbon beam energies ;
    \item \textbf{CNAO} hadrontherapy center (Pavia, Italy) \cite{Pullia_2016}, which offers advanced hadron therapy treatments using carbon ions and protons ;
    \item the \textbf{LISE} spectrometer \cite{Anne1992} of the GANIL facility (Caen, France), producing lithium ion beam of 63.6~MeV/u and carbon ion beam of 75.4~MeV/u, which can be degraded by PMMA wheel to lower energies.
    
\end{itemize}

\begin{figure*}
\begin{center}
\includegraphics[scale=0.57]{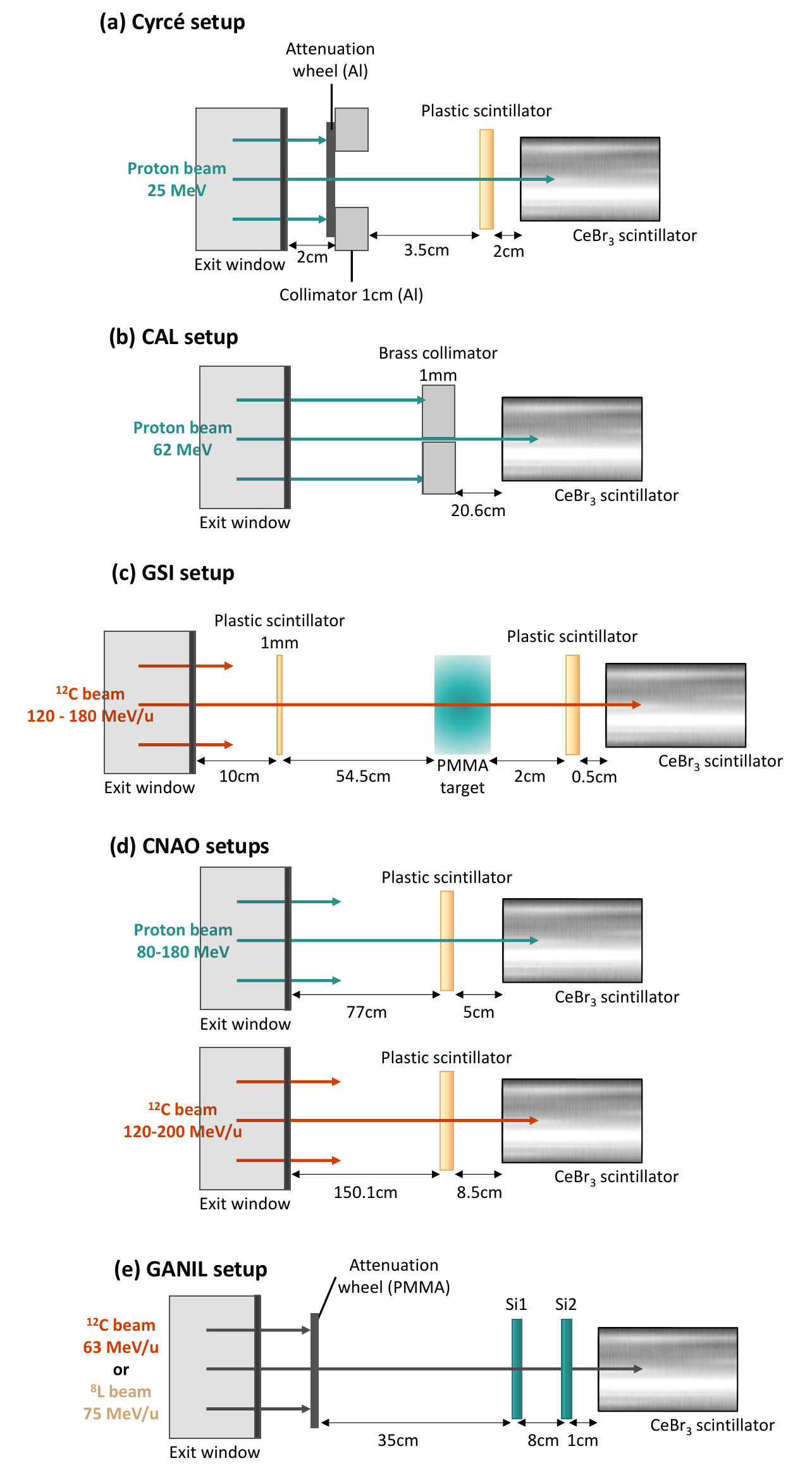}
\caption{\label{fig:setups} Schemes of the experimental setups used at Cyrcé (a), CAL (b), GSI (c), CNAO (d) and GANIL (e) facilities.}
\end{center}
\end{figure*}

The characteristics of the beams which were used are summarised in Table \ref{tab:beams}. The different experimental setups for each facilities are presented on Figure \ref{fig:setups}. 

\begin{table*}[!hbt]
\begin{center}
	\begin{tabular}{l|c|c|c}
	\hline
    \hline
	 		Facility & Ion type	& Degrador & Energy range  \\ 
 		\hline 
        \hline
			 Cyrc\'e & $^1$H & aluminum from 0.147~mm to 1.424~mm & 16 - 25~MeV \\
			CAL & $^1$H & none & 61~MeV \\
			GSI & $^{12}$C & PMMA & 120 - 180~MeV/u \\
            CNAO & $^{1}$H & none & 80 - 180~MeV \\
			CNAO & $^{12}$C & none & 120 - 200~MeV/u \\
            GANIL & $^{12}$C & PMMA wheel & 30 - 75~MeV/u \\
            GANIL & $^{8}$Li & PMMA wheel & 45 - 64~MeV/u \\
		\hline
        \hline
	\end{tabular}
	\caption{\label{tab:beams}Beam types and energies that were used for the calibration of the CeBr$_3$ and plastic scintillators.}
\end{center}
\end{table*}

In the experimental setup at GSI, a plastic scintillator of 1~mm was placed in front of the beam for another experiment which was carried out at the same beam time. At GANIL, to degrade the energy of the beam, a specific experimental setup with a wheel was developed and build, with thirty different PMMA thicknesses in order to achieve carbon beams from 30~MeV/u to 64~MeV/u and lithium beams from 45~MeV/u to 75~MeV/u. The plastic detector was replaced with two silicon detectors, each 0.2~mm thick, for an additional test phase aimed at improving resolution for the CLINM project.

\subsection{Analysis of the data}
The used analysis software package was an extension of the QAPIVI software tool \cite{Finck}, called STIVI (Software for Tracking of Ions and Vertex Imaging). The baseline refers to the initial portion of the waveform where no significant signal is expected, representing the underlying noise level of the system. The baseline was calculated by averaging the first 10 time bins of the signal, corresponding to a pre-signal region where no event is expected. The standard deviation of this interval was taken as the noise level. The determination of the signal’s rise and fall involved setting thresholds relative to the baseline, typically defined as the baseline minus 10 times the noise level. 

The integration range was defined as the interval between the rise and fall times of the signal, while the amplitude of the signal was identified as its maximum value. For each measurement condition, the distribution of signal amplitudes (in mV) were recorded over multiple events. These distributions were then fitted with Gaussian functions to extract the average amplitude — used for the detector calibration — and the standard deviation, which provides information on the signal resolution. \\

The analysis of the signal characteristics, including rise and fall times, amplitude, and charge integration, is crucial for accurately interpreting the detector response. A key aspect of this calibration process is accounting for the specific scintillation response of the detector to different particle types and their energy deposition patterns.

Initially formulated by J.B. Birk (1951), Birk’s law describes the non-linear response of scintillators to ionizing radiation, taking into accounts the scintillation quenching, a phenomenon wherein scintillator efficiency decreases as the local ionization density along the particle track increases. The classical formulation of Birk’s law \cite{Birk} is as follows:
\begin{equation}
    \frac{dL}{dx} = \frac{S \cdot \frac{dE}{dx}}{1 + k_B \cdot \frac{dE}{dx}}
\end{equation}

where \( \frac{dL}{dx} \) is the light output per unit length, \( \frac{dE}{dx} \) is the energy deposited per unit length by the particle, \( S \) is a proportionality constant that translates the energy deposited into light output (which can be attributed to scintillation efficiency), \( k_B \) is Birk’s constant, representing the scintillator-specific quenching parameter, with typical values ranging from \( 10^{-2} \) to \( 10^{-3} \) cm/MeV for common scintillators \cite{knoll2010}.

In practical applications, the differential energy loss \( \frac{dE}{dx} \) is often unknown. Instead, the total energy deposited is typically measured, \( E \), alongside the resulting pulse amplitude from the scintillator. To address this, Birk’s law was adapted to use total deposited energy rather than the differential loss. This modified approach adjusts the original constants, replacing \( S \) and \( K_B \) with \( S' \) and \( k_B' \) to reflect total energy considerations, leading to the adapted formulation:
\begin{equation}
    A = \frac{S' \cdot E + A_0}{1 + k_B' \cdot E}
\end{equation}

where \( A \) is the amplitude of the light signal measured (in mV), \( E \) is the total energy deposited (in MeV), \( S' \) is a conversion factor between the deposited energy and the amplitude, \( k_B' \) is a modified quenching parameter adapted to total energy (in MeV\(^{-1}\)), \( A_0 \) represents a pedestal or baseline signal (in mV) accounting for residual signal contributions, such as baseline shifts, electronic noise, or constant components of the detector response. While the physical expectation for ideal systems would be $A_0 = 0$, practical implementations often lead to small but non-zero pedestal values, which are accounted for in the fitting procedure.
\\

The adjusted conversion factor \( S' \) serves as a proportionality factor to convert the deposited energy \( E \) into the expected amplitude \( A \), accounting for the intrinsic scintillation light yield as well as detector-specific effects such as photon detection efficiency, photomultiplier gain, and signal shaping, with typical values ranging from \( \sim 10^{-2} \) to \( 10^0 \) mV.MeV\(^{-1}\) \cite{knoll2010}\cite{Birk}\cite{bross2007}. Unlike \( S \), which relates to differential energy loss, \( S' \) is calibrated to reflect the detector’s pulse response to total energy deposition. The adjusted quenching parameter \( k_B' \) represents saturation effects related to total energy deposited, with values typically ranging from \( 10^{-2} \) to \( 10^{-3} \, \text{MeV}^{-1} \) for inorganic scintillators \cite{Cecil} and from \( 10^{-3} \) to \( 10^{-4} \ \text{MeV}^{-1} \) for organic or plastic scintillators \cite{Brooks}, aligning with Birk’s quenching model. The pedestal \( A_0 \) accounts for baseline signal amplitude from residual detector response, typically in the low mV range. The adapted Birk’s law maintains the quenching effect via \( k_B' \) while incorporating linear responses at low energy deposition through \( S' \). It is important to emphasize that this formulation of Birks' law is used here as a phenomenological model to effectively parametrize the nonlinear response of the detection system. The parameters $S'$, $k_B'$, and $A_0$ are not intrinsic properties of the scintillating material alone, but rather effective parameters that depend on the full detection chain, including photomultiplier gain, applied voltage, and signal processing conditions. Therefore, the values obtained in this work should not be interpreted as fundamental characteristics of CeBr$_3$, but rather as calibration parameters valid under specific experimental conditions. These three parameters, including the pedestal $A_0$, are dependent on the geometry of the scintillators and the high voltage applied to the PMT. Thus, they were considered as free parameters in the fit of the detector response to ions, and represent the calibration parameters values of our detectors. \\

% For the next sections, the function will be used as follows, with $p_0$, $p_1$ and $p_2$ the variable parameters, and, for simplicity, \( S' \) and \( k_B' \) will be refereed to as  \( S \) and \( k_B \), respectively:
% \begin{equation}
%     A=\frac{p_0*E+p_1}{1+p_2*E}
% \end{equation}
% These three parameters, including the pedestal $p_0$, are dependent on the geometry of the scintillators and the high voltage applied to the PMT. Thus, they were considered as free parameters in the fit of the detector response to ions, and represent the calibration parameters values of our detectors.

Additionally, the optimization of the applied voltage was investigated to determine the best operating conditions for different ion species, ensuring both signal quality and accurate particle identification.

Beyond energy calibration, another crucial aspect of the system's characterization is its suitability for time-of-flight (TOF) measurements. The ability to resolve detection times with high precision is fundamental for identifying secondary particles and reconstructing their kinematics. Therefore, the time resolution of the telescope was assessed by quantifying detection time differences and analyzing the distribution’s standard deviation, providing key insights into the temporal capabilities of the system.

\subsection{Monte Carlo simulations}
The Monte Carlo code \textsc{Geant}4 10.07 \cite{Agostinelli2003} with the INCL++ physic list \cite{Geant4} was used to evaluate the deposited energy in the plastic scintillator, and the energy of the ions reaching the CeBr$_3$ crystal, after its entrance window. The INCL++ physics list was chosen for its proven accuracy in simulating nuclear fragmentation of light ions in the 100–400~MeV/u range, which is highly relevant to hadrontherapy scenarios.\\ 

% While it is not necessary to take into account the thickness of this entrance window for $\gamma$-rays detection, it needs to be considered for ions detection, as they will lose a non-negligible amount of energy inside. 

Simulations were also used to evaluate the energy straggling and beam scattering. Indeed, contrary to gamma-rays measurements, where the standard deviation around the mean energy value corresponds to the detector resolution at the considered energy, charged particles encounter important straggling in the different materials they cross, manifesting as a spread in the energy loss distribution, contributing to the observed broadening of the energy spectrum in detectors. Alongside, beam scattering, arising from multiple Coulomb scattering events experienced by charged particles, induces deviations from their initial trajectories, thereby influencing the spatial distribution of energy deposition.

To rigorously quantify the intrinsic energy resolution of each detector, it is imperative to disentangle the contributions from energy straggling and beam scattering from the overall detector resolution. This requires subtracting the standard deviation derived from the Monte Carlo simulations, $\sigma_{G4}$, from the standard deviation obtained from the experimental data calibration, $\sigma_{det}$, as per the equation \cite{Schardt2010}\cite{salvador2020}:
\begin{equation}
    \sigma_E=\sqrt{\sigma_{det}^2-\sigma_{G4}^2}
\end{equation}
where $\sigma_{det}$ corresponds to the standard deviation of the energy distribution obtained after calibration of the raw data distributions, and $\sigma_{G4}$ is the standard deviation of the energy distribution predicted by Geant4. This deviation does not take into account any detector response, and therefore corresponds only to the straggling and beam scattering.\\

Moreover, the fitting of detector resolution as a function of collected energy using the proposed equation facilitates a comprehensive characterization of the resolution behavior across varying energy depositions. The parameters a and b in the equation (3) capture the nuanced dependence of resolution on energy deposition \cite{leo1994}\cite{hampel1990}:
\begin{equation}
    \frac{\sigma_E}{E}=a+\frac{b}{\sqrt{E}}
\end{equation}
with a and b being free parameters for the fit.

\section{Results}
\label{Results}
\subsection{Response of the plastic scintillator}
\begin{figure*}[!hbt]
\begin{center}
\includegraphics[scale=0.6]{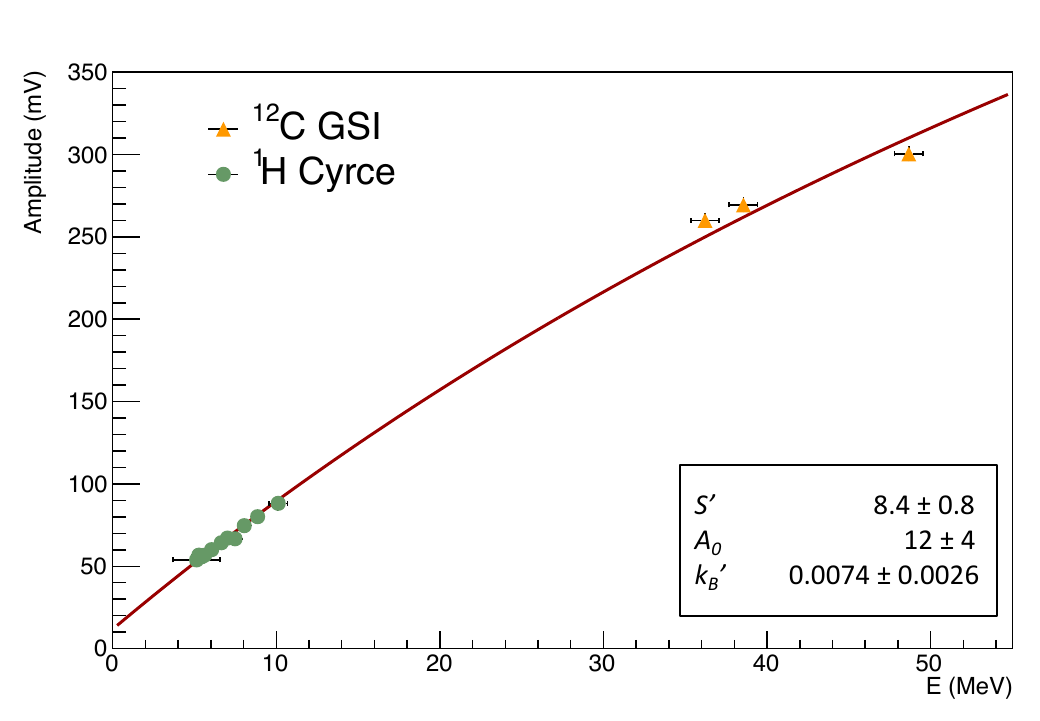}
\caption{\label{fig:calibPlastic1} Calibration curve of the plastic scintillator with the R7057 PMT at -~1100~V, with $^{12}$C beams of GSI and $^{1}$H beams of Cyrcé. Energies can be found on Table \ref{tab:beams}.}
\end{center}
\end{figure*}

\begin{figure*}
\begin{center}
\includegraphics[scale=0.65]{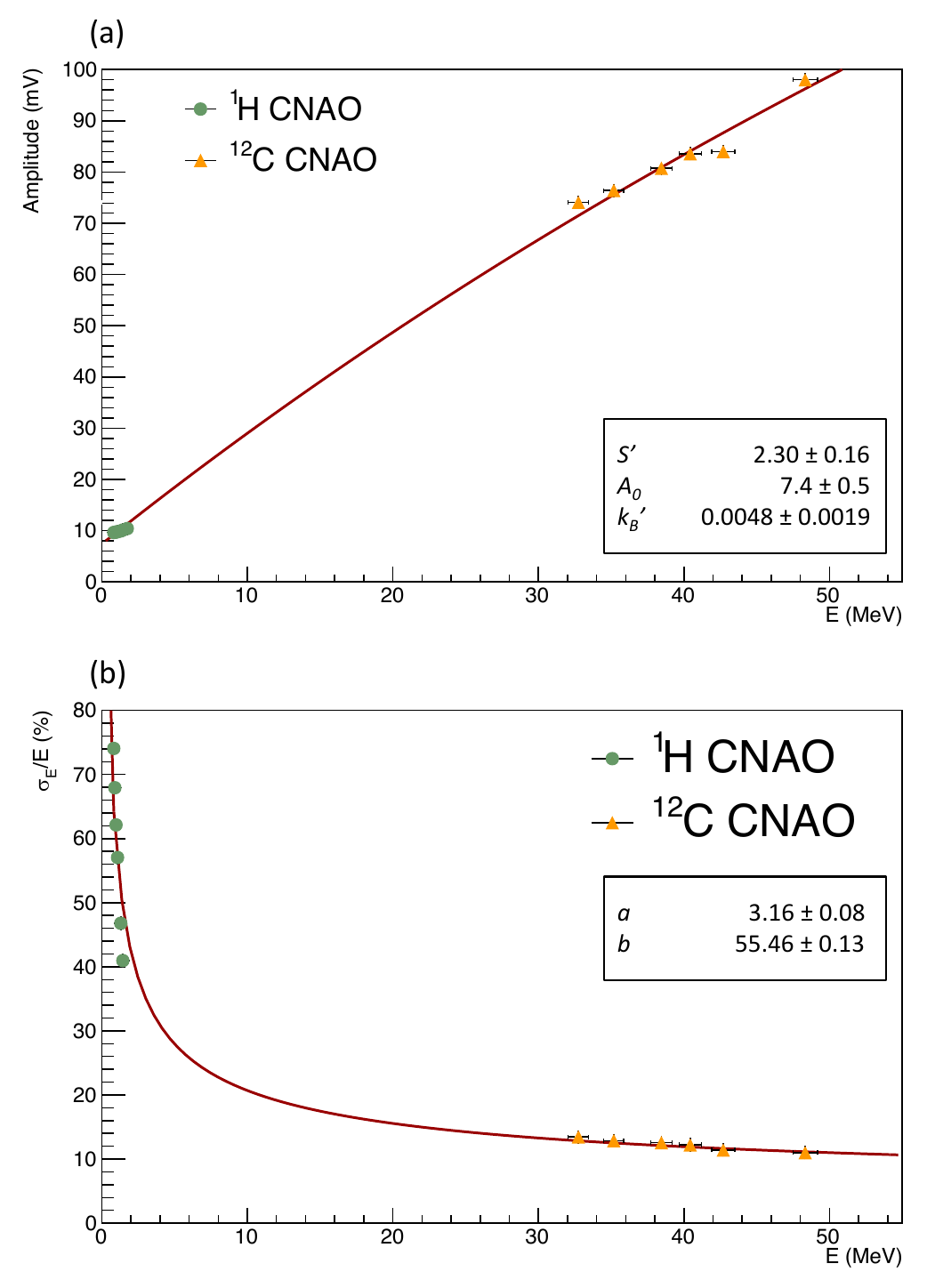}
\caption{\label{fig:calibPlastic2} Calibration curve of the plastic scintillator with the XP3990 PMT at -~1200~V (a) and energy resolution as a function of the deposited energy (b), with $^{12}$C and $^{1}$H beams of CNAO. Energies can be found on Table \ref{tab:beams}.}
\end{center}
\end{figure*}

The calibration curve of the plastic scintillator obtained with both protons and $^{12}$C ions is presented on Figure \ref{fig:calibPlastic1} and \ref{fig:calibPlastic2}(a) for both photomultipliers. 
It can be observed in both cases that the Birk's law is verified up to 50~MeV of deposited energy, which corresponds to the energy deposited by $^{12}$C ions of 180~MeV/u (after 30~mm of PMMA) for Figure \ref{fig:calibPlastic1} and by $^{12}$C ions of 120~MeV/u for Figure \ref{fig:calibPlastic2}(a). The two calibrations constants cannot be compared with each other because of the two different PMT and voltage applied (-~1200~V and -~1100~V). 

The energy resolution of the scintillator with the XP3990 PMT is also shown, as a function of the deposited energy, on Figure \ref{fig:calibPlastic2}(b). This analysis takes into account the contribution of beam straggling and scattering, as evaluated by Geant4 simulations. However, some experimental uncertainties such as beam energy spread, angular divergence, or alignment tolerances are not fully modeled and may introduce additional broadening in the measured spectra. These effects can contribute to a slight overestimation of the detector’s intrinsic resolution. The uncertainty in energy loss can be in the order of several MeV, depending on the thickness and composition of the material. Similarly, scattering can cause angular deviations up to
a few degrees, impacting the spatial resolution of the detector.

\subsection{Response of the CeBr$_3$ to ions}
\begin{figure*}
\begin{center}
\includegraphics[scale=0.7]{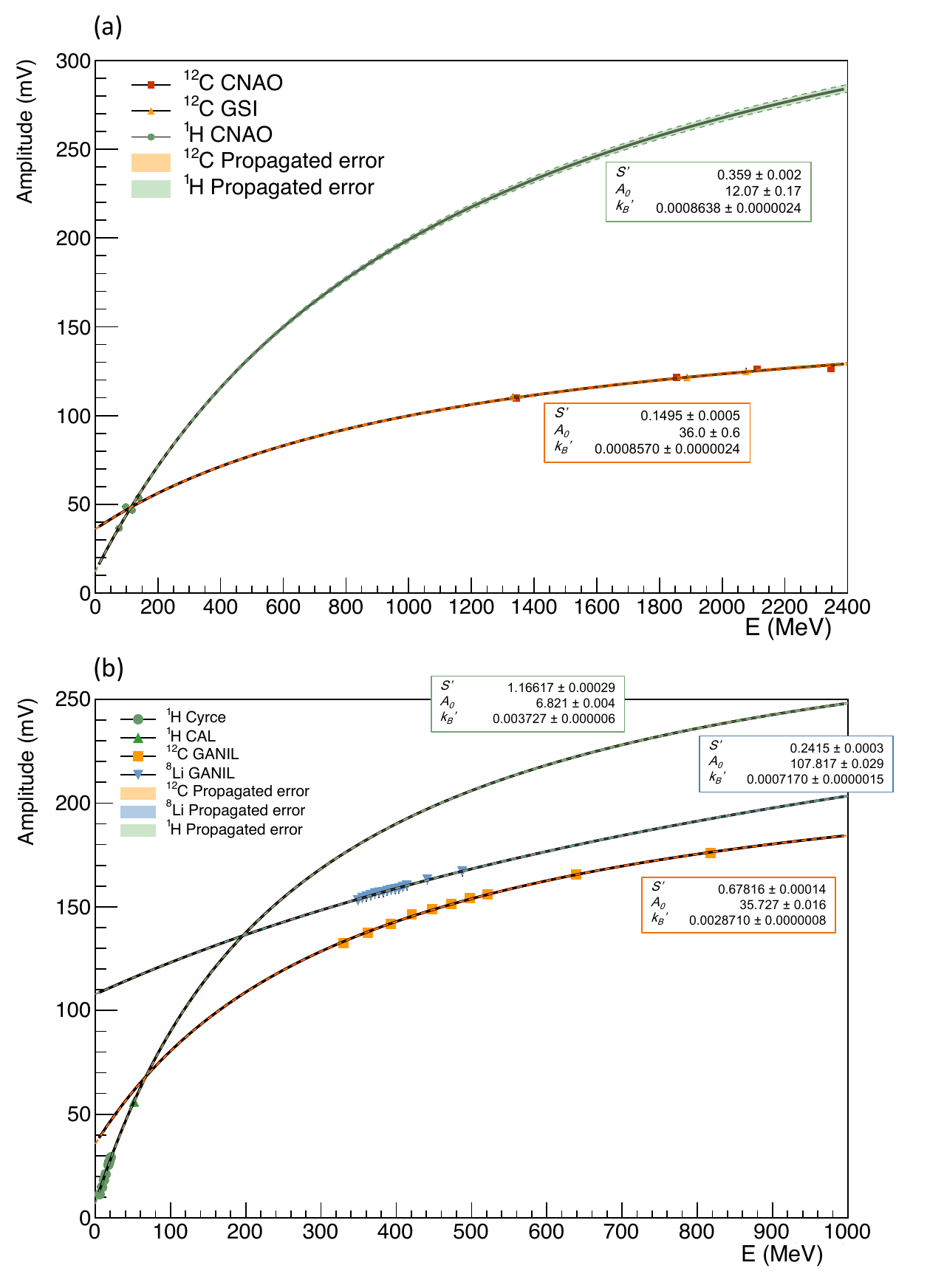}
\caption{\label{fig:calibCeBr} Calibration curves of the CeBr$_3$ crystal scintillator obtained with both protons and $^{12}$C, with an applied voltage of +~350~V for (a) and obtained with protons, $^8$Li and $^{12}$C, with an applied voltage of +~400~V for (b). Energies can be found on Table \ref{tab:beams}. The solid lines correspond to fits using Birks’ law, and the shaded bands represent the propagated uncertainties from the fit parameters.}
\end{center}
\end{figure*}

On Figure \ref{fig:calibCeBr}, the calibration curves of the CeBr$_3$ crystal scintillator obtained with both protons and $^{12}$C, with an applied voltage of +~350~V for (a) and obtained with protons, $^8$Li and $^{12}$C, with an applied voltage of +~400~V for (b), are presented.
For each ion type (carbon, lithium and proton), distinct calibration functions are applied and extrapolated to energy ranges not measured. This differentiation is consistent with the findings in \cite{Miller_2018}, where the scintillators’ light output showed varying responses based on the isotope.
Remarkably, the Birk’s constant appears similar in order of magnitude and numerical value for all ions, reflecting a robust consistency in the calibration results. Adherence to Birk’s law is observed up to 2350~MeV of deposited energy, corresponding to the energy deposited by $^{12}$C of 200~MeV/u (after traversing 2~mm of PMMA). Notably, the plotted points for protons, lithium and carbon ions exhibit a conformal distribution, suggesting a consistent response of the scintillator across different particle types and energies.

All the fit parameters values collected for both plastic and CeBr$_3$ and for all the voltage used can be found in Table \ref{tab:parBirks}. As mentioned in section \ref{MatMeth}, the values of \( k_B' \) and \( S' \) obtained here are effective and reflect not only the quenching behavior of the scintillator but also the influence of experimental parameters such as the applied high voltage and PMT gain. The variation in the \( k_B' \) values observed across different configurations (e.g., +350~V vs. +400~V) is therefore expected and does not indicate a contradiction with the underlying physics.
It can be observed that, depending on the voltage and the photomultiplier gain, distinct Birk's constants are obtained. 

Although the absolute values of \( k_B' \) vary between ions and voltage settings, they remain within the expected order of magnitude (10$^{-3}$–10$^{-2}$~MeV$^{-1}$) reported in the literature for inorganic scintillators. As these constants are effective and depend on experimental parameters, perfect agreement between all conditions is not expected. Instead, the observed consistency in trends and ranges supports the validity of the calibration model.
For all detector configurations, the pedestal was found to be close to zero, confirming that the baseline subtraction was effective.

\begin{table*}[!hbt]
\begin{center}
	\begin{tabular}{|l|c|c|l}
	\hline
	 		Detector & \( S' \)	& \( k_B' \) (MeV$^{-1}$) \\ 
 		\hline 
        \hline
			Plastic -1100V & 8.4 $\pm$ 0.8 & 0.0074 $\pm$ 0.0026 \\
            Plastic -1200V & 2.30 $\pm$ 0.16 & 0.0048 $\pm$ 0.0019\\
			CeBr$_3$ +350V (Carbons) & 0.1495 $\pm$ 0.0005 & 0.0008570 $\pm$ 0.0000024 \\
            CeBr$_3$ +350V (Protons) & 0.3590 $\pm$ 0.0002 & 0.0008638 $\pm$ 0.0000024 \\
			CeBr$_3$ +400V (Carbons) &  0.67816 $\pm$ 0.00014 & 0.0028710 $\pm$ 0.0000008 \\
            CeBr$_3$ +400V (Protons) & 1.16617 $\pm$ 0.00029 & 0.003727 $\pm$ 0.000006 \\
            CeBr$_3$ +400V (Lithiums) & 0.2415 $\pm$ 0.0003 & 0.0007170 $\pm$ 0.0000015 \\
		\hline
	\end{tabular}
	\caption{\label{tab:parBirks}Obtained values of free parameters from Birks' law used to fit the detectors calibration curves.}
\end{center}
\end{table*}

\begin{figure*}
\begin{center}
\includegraphics[scale=1.2]{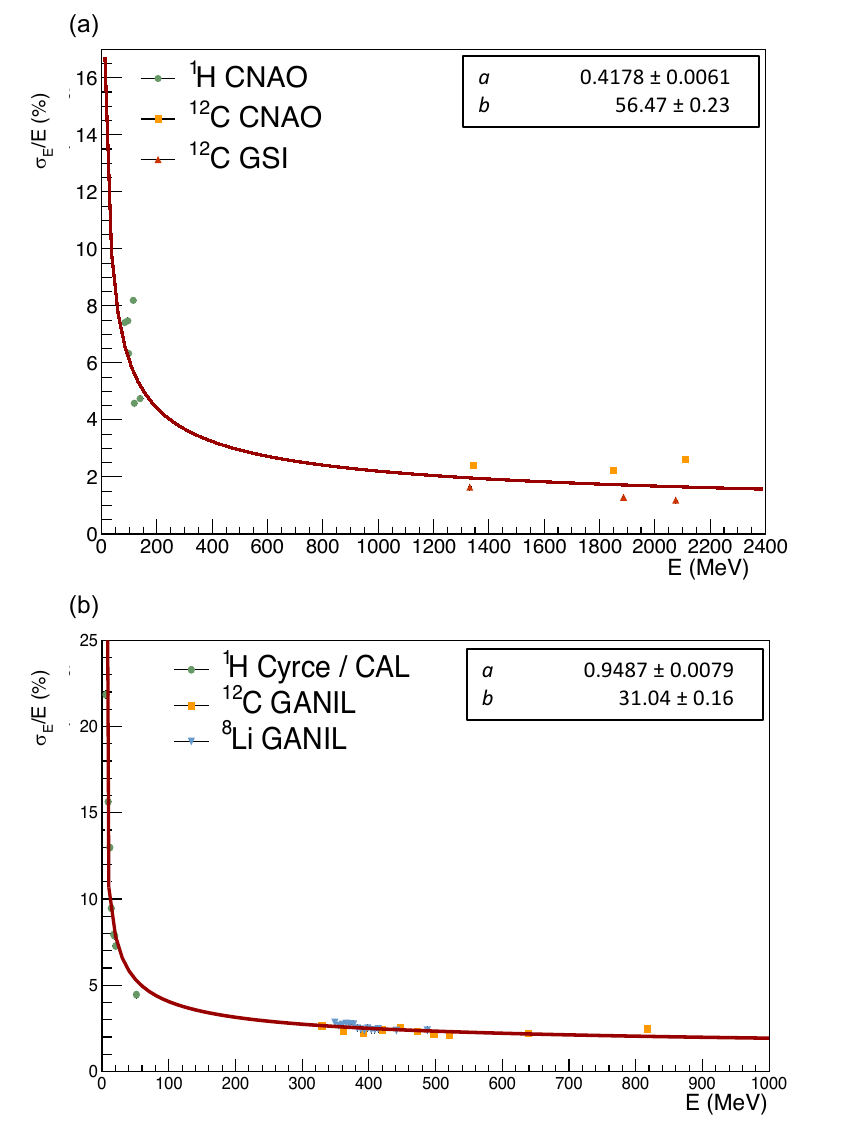}
\caption{\label{fig:resolution}Resolution curves of the CeBr$_3$ crystal scintillator obtained with both protons and $^{12}$C, with an applied voltage of +~350~V for (a) and obtained with protons, $^8$Li and $^{12}$C, with an applied voltage of +~400~V for (b). Energies can be found on Table \ref{tab:beams}. }
\end{center}
\end{figure*}

On Figure \ref{fig:resolution} are presented the energy resolution distribution of the CeBr$_3$ obtained with both protons and $^{12}$C for a +~350~V voltage applied for (a), and of the CeBr$_3$ obtained with protons, $^8$Li and $^{12}$C for a +~400~V voltage applied for (b). Remarkably, a resolution in the order of maximum 10~MeV is observed for all ion types, underscoring the precision of the crystal in measuring energy depositions.

\begin{figure*}[!hbt]
\begin{center}
\includegraphics[scale=0.9]{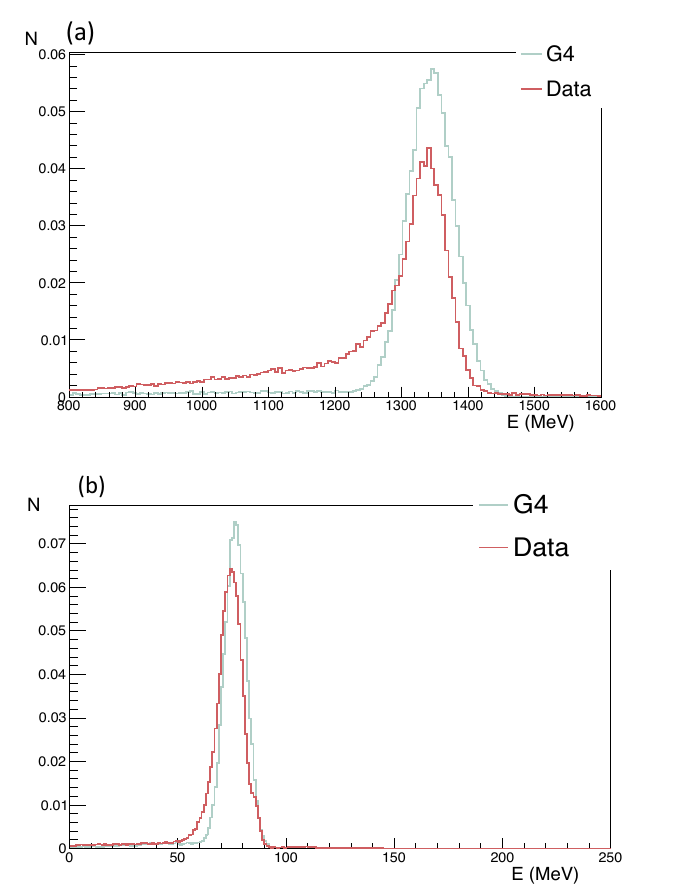}
\caption{\label{fig:Calib_test} CeBr$_3$ deposited energy comparison between G4 simulation and calibrated normalized data with a 120~MeV/u $^{12}$C beam (a) and a 100~MeV $^1$H beam.}
\end{center}
\end{figure*}

To validate the calibration of the CeBr$_3$ scintillator, experimental data in energy are compared with Geant4 simulations, as depicted in Figure \ref{fig:Calib_test}. This comparison offers insights into the reliability of the methodology across proton, lithium and carbon ion measurements. Looking into the comparison of deposited energy in CeBr$_3$ between Geant4 simulations and calibrated experimental data, using a 120~MeV/u $^{12}$C beam as illustrated in Figure \ref{fig:Calib_test}(a), the presence of a fragmentation tail is revealed in the experimental data, but less present in the simulation. However, the simulation accurately reproduces the energy peak distribution. Similarly, Figure \ref{fig:Calib_test}(b) depicts the comparison of CeBr$_3$ deposited energy between Geant4 simulations and calibrated experimental data using a 80~MeV $^{1}$H beam. Here, a discrepancy of less than 5~MeV in the energy peak can be observed. While this deviation is noticeable, it falls within an acceptable range for ensuring the reliability of our measurements. This range, typically defined as ±~5~ $\%$, accounts for minor variations in experimental conditions and calibration inconsistencies. Specifically, a 5~MeV discrepancy in the context of a 80~MeV $^{1}$H beam represents a deviation of about 6$\%$, which, while significant, remains manageable within the broader scope of experimental nuclear physics, where uncertainties can often span up to 10-20~$\%$ depending on the complexity of the setup and the type of particles measured \cite{hamm2012industrial}. 
%Moreover, radiolysis effects do not significantly change with slight variations in energy, maintaining relative consistency in the CLINM project context even with differences of a few percent \cite{buxton1988rate}.

\subsection{Telescope ions identification}
\begin{figure*}
\begin{center}
\includegraphics[scale=1.]{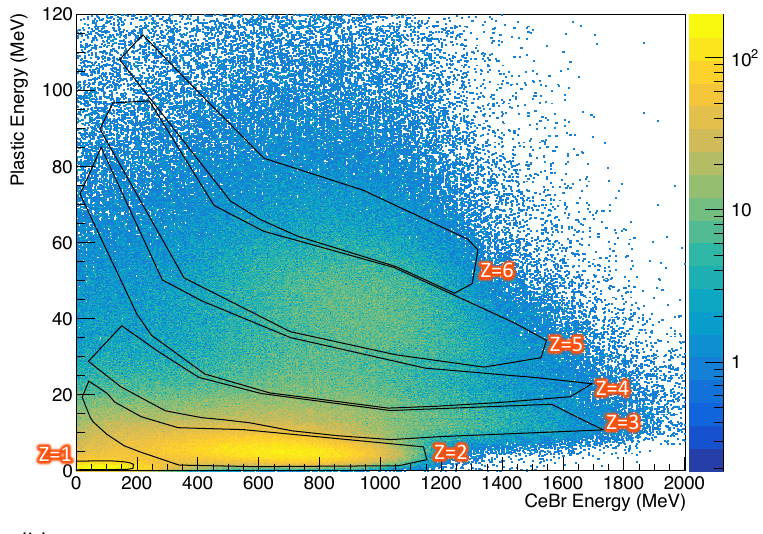}
\caption{\label{fig:DeltaEE}Energy deposited inside the plastic scintillator ($\Delta E$) as a function of the energy deposited inside the CeBr$_3$ ($E$) for a $^{12}$C beam of 200~MeV/u on a 5 cm RW3 target and the detectors at 5$^\circ$ with the Z ions discrimination showed. }
\end{center}
\end{figure*}

The identification of different ions using the $\Delta E-E$ telescope leverages the energy deposition characteristics in the plastic and CeBr$_3$ scintillators. Figure \ref{fig:DeltaEE} demonstrates the ability of the $\Delta E-E$ telescope to distinguish ions based on their charge state (Z). The plot shows the energy deposited in the plastic scintillator ($\Delta E$) as a function of the energy deposited in the CeBr$_3$ scintillator ($E$), for the secondary particles produced by a 200~MeV/u $^{12}C$ ion beam on a RW3 target (tissue equivalent), revealing distinct structures. This figure is intended as a qualitative illustration of the telescope’s ability to discriminate charged fragments by charge (Z) using $\Delta E-E$ correlations. A more detailed analysis, including Monte Carlo comparison, separation efficiency, and species contamination rates, is beyond the scope of this work and will be presented in a dedicated publication.

Each ion species (Z = 1 to Z = 6) forms a distinct branch in the $\Delta E - E$ space, reflecting its unique energy loss and remaining energy characteristics. The contours in the Figure are intended as visual guides based directly on the observed data, to emphasize visible patterns in the distributions. This demonstrates the ability of the $\Delta E-E$ telescope to identify secondary charged particles produced by therapeutic ion beams interacting with tissue-equivalent targets under experimental conditions representative of clinical hadrontherapy.

\subsection{Time performance}
\begin{figure*}
\begin{center}
\includegraphics[scale=0.7]{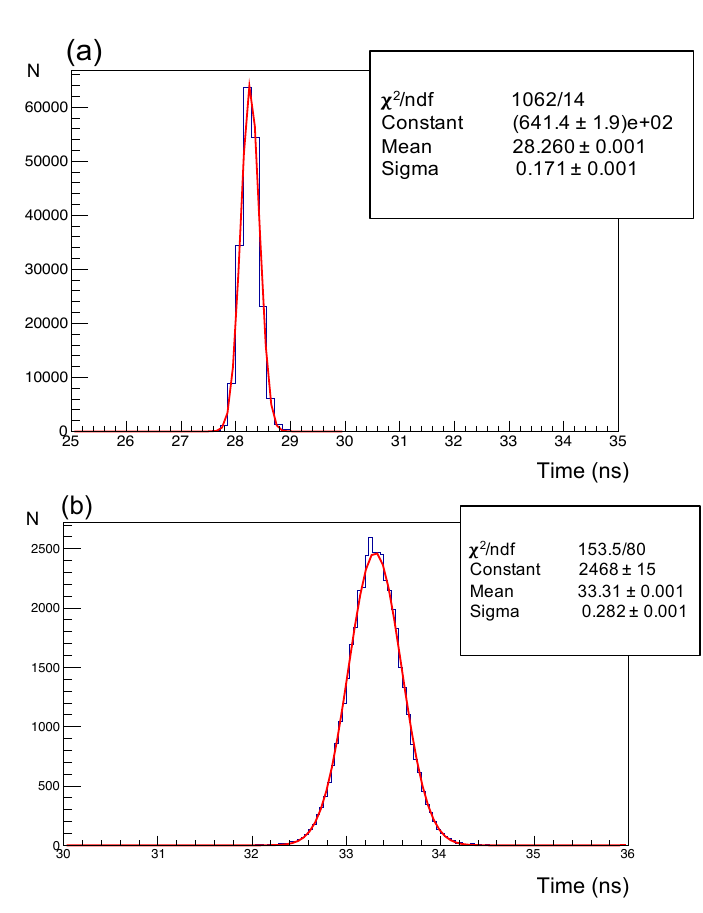}
\caption{\label{fig:time180} Resolution in time (measured time difference) between the plastic with the CeBr$_3$ and R7057 PMT with a 25~MeV $^1$H beam (a) and XP3990 PMT with a 180~MeV/u $^{12}$C beam (b)}
\end{center}
\end{figure*}

A good time resolution ensures that the two measurements, energy loss and total energy, can be accurately correlated in time, aiding in precise particle identification. Moreover, this test is necessary as the telescope will also be used in $\Delta E$-ToF measurements for high-energy ions, as well as for detecting gammas and neutrons, necessitating rigorous temporal performance assessments. For instance, the FRACAS experiment achieved a time resolution of around 300~ps for similar high-energy ion beams \cite{salvador2020timing}. These benchmarks illustrate the standards required for effective particle detection. The time resolution values were extracted from the experiments performed, with both protons and $^{12}$C. On Figure \ref{fig:time180}, the time discrepancies between the two detector are presented, for a 25~MeV protons beam (a) and a 180~MeV/u $^{12}$C beam (b). 

The temporal resolution obtained from those distributions for the 180~MeV/u $^{12}$C beam, is of 282 $\pm$ 1~ps. Similarly, a time resolution of 171 $\pm$ 1~ps was measured between the plastic and CeBr$_3$ detectors using a 25~MeV proton beam.

The uncertainties on the time resolution correspond to statistical errors only. The increase of the time resolution value for carbon ions is mainly due to the low voltage that was applied to the cerium bromide PMT (+~350~V). These values represent crucial benchmarks in assessing the time performance of the detection system across varying beam energies and particle types. Such temporal resolution demonstrates the system’s potential for future implementation of $\Delta E$–ToF particle identification methods, and enhances the overall capabilities of the experimental setup for the CLINM purpose.

\section{Discussion}
\label{Discussion}
The calibration of the plastic scintillator yielded a calibration curve with two different PMTs both adhering to Birks' law, up to 50~MeV of deposited energy. This result demonstrates the reliability of the plastic scintillator's response to ion beams within this energy range. The energy resolution of the plastic scintillator was also examined, and it is noteworthy that the resolution improves as the deposited energy increases. This is consistent with the expected behavior, as higher energy ions generate more scintillation light, resulting in better energy resolution.

The CeBr$_3$ crystal scintillator was calibrated with proton, $^{8}$Li and $^{12}$C ion beams. The calibration results show that the crystal scintillator follows the Birks' law up to 2350~MeV of deposited energy, which corresponds to the energy deposited by 200~MeV/u $^{12}$C ions (after traversing 2~mm of PMMA). This extended reliability is a valuable characteristic, as it allows for the accurate measurement of a wide range of ion energies encountered in hadrontherapy treatment (80 - 200~MeV/u for $^{1}$H beam and 120 - 400~MeV/u for $^{12}$C beam). \\
Furthermore, examination of calibration curves obtained with protons at different voltages, +~350~V and +~400~V exhibited consistency, indicative of the CeBr$_3$ scintillator’s efficient response to varied voltage settings for ion detection. 

Notably, the Birks’ law reflected a quenching effect of the scintillators, with an increase in voltage corresponding to an increase in the Birks constant. This observation underscores the adaptability of the scintillator to diverse experimental conditions and provides options for optimizing its performance. The adherence to Birks’ law across multiple ion types (protons, $^{8}$Li, and $^{12}$C) and energy ranges (16~MeV to 2350~MeV deposited energy) is a notable result, as it ensures the system’s reliability for clinical applications. This level of precision has not been demonstrated previously for CeBr$_3$ scintillators in such conditions.

In addition to energy resolution, the time performance of our $\Delta E-E$ telescope was also evaluated. Precise time measurements enable the correlation of energy loss and total energy, facilitating particle identification and background rejection. By precisely calibrating and characterizing the $\Delta E-E$ telescope, this study establishes a robust foundation for future measurements of secondary charged particles. Such data, once acquired in tissue-equivalent material, will provide valuable input for Monte Carlo benchmarking and, in the long term, may help improve the accuracy of treatment planning systems.

The time resolution measurements exhibit a resolution of 282 $\pm$ 1~ps for $^{12}$C ions of 180~MeV/u and 171 $\pm$ 1~ps for 25~MeV protons. These results underscore the capability of this telescope to provide precise measurements under experimental conditions that are representative of clinical hadrontherapy scenarios. The narrower time resolution distribution for protons compared to $^{12}$C ions can be attributed to the differences in their masses and velocities. The timing resolution of 171–282~ps, coupled with an energy resolution on the order of 10~MeV, represents a substantial advancement in the detection of secondary particles. This precision enables not only better particle identification but also provides accurate input for Monte Carlo models, thereby improving the reliability of dose calculations.

The precise particle identification offered by the $\Delta E-E$ telescope is critical for characterizing the secondary particles in clinical beams. Such measurements can improve our understanding of fragmentation processes in tissues and provide accurate inputs for Monte Carlo simulations, leading to more reliable treatment planning. The system's capability to detect ions across a wide range of charge states ensures its applicability in various clinical scenarios.

It should be emphasized again that Birks' law was used here as an effective parametrization of the system response. The extracted constants are not intrinsic to the CeBr$_3$ material itself, but reflect the global response of the detector system under specific operating conditions. This empirical modeling approach is particularly suited to experimental configurations where full analytical modeling of light yield and collection is impractical.

\section{Conclusion}
In the context of hadrontherapy, precision in measuring energy deposition stands as a fundamental requirement for optimizing treatment plans. In this work, the calibration and performance evaluation of a $\Delta E-E$ telescope was presented, designed for the detection of secondary charged particles generated by ion beams under experimental conditions representative of clinical hadrontherapy. This telescope combines a thin plastic scintillator and a CeBr$_3$ crystal scintillator, optimized for precision in energy and time measurements.

The calibration results demonstrate adherence to Birks' law, with the plastic scintillator effectively detecting energy deposition up to 50~MeV and up to 2350~MeV for the the CeBr$_3$ crystal. These calibrations were achieved under multiple voltage settings and across a range of particle types (protons, $^{8}$Li, $^{12}$C), showcasing the versatility of the detection system.

The energy resolution was determined to be on the order of 10~MeV for both scintillators, ensuring precise identification of secondary particles. The system's time resolution was measured at 282 $\pm$ 1~ps for 180 MeV/u $^{12}$C ions and 171 $\pm$ 1 ps for 25~MeV protons. These results confirm the potential of the telescope as a reliable system for $\Delta E-E$ and $\Delta E$-ToF measurements, capable of separating secondary particles generated during beam-tissue interactions.

The telescope was tested under experimental conditions designed to replicate the physical characteristics of clinical hadrontherapy, and was shown to perform effectively within the relevant energy ranges and ion species. The system is not intended for direct use on patients, but rather for producing precise and reliable nuclear data using tissue-equivalent targets. These measurements provide essential input for Monte Carlo simulations, ultimately improving the accuracy of treatment planning systems.

\acknowledgments

Authors want to thank Claire-Anne Reidel and Christoph Schuy from the GSI biophysics department, whose help was crucial for carrying out this work. We also want to thank Angelica Facoetti from CNAO, Petter Hofverberg from CAL facility, and Michel Pellicioli and Jacky Schuler from the Cyrc\'e facility. Finally, this work would also not have been possible without the valuable help of C\'edric Mathieu and Thomas Adam. This work was possible thanks to the support of the HITRI+ project, the funding of the European Union’s Horizon 2020 Research and innovation (Grant Agreement Nº101008548), and the ANR project funding ANR-23-CE31-0014.

\end{document}